\documentclass{mn2e}

\usepackage{amsmath}
\usepackage{amssymb}
\usepackage{graphicx,epsfig}

\def\gsim{\;\lower4pt\hbox{${\buildrel\displaystyle >\over\sim}$}\;}
\def\lsim{\;\lower4pt\hbox{${\buildrel\displaystyle <\over\sim}$}\;}
\def\grls{\;\lower4pt\hbox{${\buildrel\displaystyle >\over <}$}\;}

\title[Two-Phase Formation of SMBHs]
{Forming supermassive black holes by\\
accreting dark and baryon matter}

\author[Hu, Shen, Lou, Zhang]
{Jian Hu$^1$, Yue Shen$^{1,5}$, Yu-Qing Lou$^{1,2,3,4}$,
and Shuangnan Zhang$^{1,6,7,8}$\\
$^1$Physics Department and Tsinghua Center for
Astrophysics (THCA), Tsinghua University, Beijing 100084, China;\\
hujian98@mails.tsinghua.edu.cn; 
yshen@astro.princeton.edu;
louyq@tsinghua.edu.cn; zhangsn@mail.tsinghua.edu.cn\\
$^2$Centre de Physique des Particules de Marseille (CPPM)
/Centre National de la Recherche Scientifique (CNRS)\\
\qquad\quad
/Institut National de Physique Nucl\'eaire et de Physique
des Particules (IN2P3) et Universit\'e \\ \qquad\ \
de la M\'editerran\'ee Aix-Marseille II,
163, Avenue de Luminy Case 902 F-13288 Marseille, Cedex 09, France;\\
$^3$Department of Astronomy and Astrophysics, The University
of Chicago, 5640 South Ellis Avenue, Chicago, IL 60637 USA;\\
$^4$National Astronomical Observatories of China, Chinese
Academy of Sciences, A20, Datun Road, Beijing 100012, China;\\
$^5$Department of Astrophysical Sciences, Peyton Hall, 
Princeton University, Princeton, NJ 08544 USA;\\
$^6$Department of Physics, University of Alabama in Huntsville,
Huntsville, AL 35899 USA;\\
$^7$National Space Science and Technology Center, 
320 Sparkman Drive, SD50, Huntsville, AL 35805 USA;\\
$^8$Institute of High Energy Physics, Chinese Academy of Sciences,
P.O. Box 918-3, Beijing 100039, China.}
\date{Accepted 2004... Received 2003...;
in original form 2003}\date{Accepted .
      Received ;
      in original form }

\begin{document}
\maketitle

\begin{abstract}
Given a large-scale mixture of self-interacting dark matter (SIDM)
particles and baryon matter distributed in the early Universe, we
advance here a two-phase accretion scenario for forming supermassive 
black holes (SMBHs) with masses around $\sim 10^9 M_{\odot}$ at high 
redshifts $z\ (\gsim 6)$. The first phase is conceived to involve a rapid 
quasi-spherical and quasi-steady Bondi accretion of mainly SIDM particles 
embedded with baryon matter onto seed black holes (BHs) created at 
redshifts $z\lsim 30$ by the first generation of massive Population III 
stars; this earlier phase rapidly gives birth to significantly enlarged 
seed BH masses of $M_{\hbox{\tiny BH},t_1}\backsimeq 1.4\times 10^6\ 
M_\odot\ \sigma_0/(1\hbox{ cm}^2\hbox{ g}^{-1})(C_s/30\hbox{ km s}^{-1})^4$ 
during $z\sim 20-15$, where $\sigma_0$ is the cross section per unit 
mass of SIDM particles and $C_s$ is the velocity dispersion in the 
SIDM halo referred to as an effective ``sound speed". The second phase 
of BH mass growth is envisaged to proceed primarily via baryon accretion, 
eventually leading to SMBH masses of $M_{\hbox{\tiny BH}}\sim 10^9\ M_\odot$;
such SMBHs may form either by $z\sim 6$ for a sustained accretion at the 
Eddington limit or later at lower $z$ for sub-Eddington mean accretion 
rates. In between these two phases, there is a transitional yet sustained 
diffusively limited accretion of SIDM particles which in an eventual 
steady state would be much lower than the accretion rates of the two 
main phases. We intend to account for the reported detections of a 
few SMBHs at early epochs, e.g., SDSS 1148+5251 and so forth, without 
necessarily resorting to either super-Eddington baryon accretion or 
very frequent BH merging processes. Only extremely massive dark SIDM 
halos associated with rare peaks of density fluctuations in the early 
Universe may harbour such early SMBHs or quasars. Observational 
consequences are discussed. During the final stage of accumulating a 
SMBH mass, violent feedback in circumnuclear environs of a galactic 
nucleus leads to the central bulge formation and gives rise to the 
familiar empirical $M_{\hbox{\tiny BH}}-\sigma_b$ correlation inferred 
for nearby normal galaxies with $\sigma_b$ being the stellar velocity 
dispersion in the galactic bulge; in our scenario, the central SMBH 
formation precedes that of the galactic bulge.
\end{abstract}


\section{Introduction}

On the basis of various observational diagnostics and numerous case 
studies, supermassive black holes (SMBHs) are now widely believed to be
ubiquitous, particularly at the nuclei of both normal and active
galaxies (e.g., Kormendy \& Richstone 1995; Haehnelt 2004). As the
central gravitational engines to power most energetic activities
of quasi-stellar objects (QSOs) or quasars (e.g., Salpeter 1964;
Lynden-Bell 1969; Bardeen 1970), SMBHs dynamically impact the
formation and evolution of host galaxies (e.g., Silk \& Rees 1998;
Page, Stevens, Mittaz \& Carrera 2001; King 2003; Murray, Quartaert
\& Thompson 2005). Their most tantalizing manifestations are the
observed $M_{\hbox{\tiny BH}}-M_{\hbox{\tiny bulge}}$ correlation
(e.g., Magorrian et al. 1998; Laor 2001; H\"aring \& Rix 2004) and
its tighter version --- the $M_{\hbox{\tiny BH}}\propto\sigma_b^{4.2}$
correlation
%M_{BH}\propto \sigma_b^{4.02}
for both active and normal galaxies (e.g., Gebhardt et al. 
2000; Ferrarese \& Merritt 2000; Tremaine et al. 2002), 
where $M_{\hbox{\tiny BH}}$ is the black hole mass, 
$M_{\hbox{\tiny bulge}}$ is the galactic central bulge mass
and $\sigma_b$ is the stellar velocity dispersion in the 
galactic bulge.

Given substantial progresses in probing and understanding the basic
physics of SMBHs as well as galaxy formation and evolution over past 
several decades (e.g., Lynden-Bell 1969), much still remain to be 
learned and explored because of the somewhat speculative nature of 
the subject to a certain extent. During the extensive Sloan Digital 
Sky Survey (SDSS), the newly reported SMBH with a mass of
$M_{\hbox{\tiny BH}}\sim 3\times 10^9\ M_\odot$ in the quasar SDSS 
1148+5251 (Fan et al. 2003; Willott et al. 2003) at a high redshift 
$z=6.43$ particularly highlights the outstanding mystery of the 
rapid BH mass growth in the early Universe and reveals inconsistency 
with the local $M_{\hbox{\tiny BH}}-\sigma_b$ relation (e.g., Walter 
et al. 2004).\footnote{In this context, we note in passing the recent 
detection of a gamma-ray burst afterglow with a high redshift $z\gsim 6$.}

For a sustained Eddington accretion of baryon matter, the mass growth 
rate $\dot{M}_{\hbox{\tiny BH}}$ of a BH is presumed to be proportional 
to the black hole mass $M_{\hbox{\tiny BH}}$, namely
\begin{equation}
\dot{M}_{\hbox{\tiny BH}}=M_{\hbox{\tiny BH}}/t_{\hbox{\tiny Sal}}\ ,
\end{equation}
where $t\hbox{\tiny Sal}$ is the so-called 
Salpeter timescale (Salpeter 1964)
$$
t_{\hbox{\tiny Sal}}\equiv\frac{\epsilon M_{\hbox{\tiny BH}}c^2}
{(1-\epsilon)L}=3.9\times10^7\hbox{ yr}
\frac{\epsilon L_{\hbox{\tiny Edd}}}{0.1(1-\epsilon)L}\
$$
with $c$, $L$, $L_{\hbox{\tiny Edd}}$ and $\epsilon$ being 
the speed of light, the luminosity, the Eddington luminosity 
and the radiative efficiency, respectively. For constant 
$\epsilon$ and $L/L_{\hbox{\tiny Edd}}$ parameters, the BH 
mass grows exponentially in the form of
\begin{equation}
M_{\hbox{\tiny BH}}(t)=M_0\exp\ [(t-t_0)/t_{\hbox{\tiny Sal}}]\ ,
\end{equation}
where $M_0$ is the seed BH mass and $t_0$ is the initial time of accretion. 
Recent observations suggest $L/L_{\hbox{\tiny Edd}}\lsim 1$ (e.g., 
Vestergaard 2004; McLure \& Dunlop 2004) and $\epsilon\gtrsim 0.1-0.2$ 
(e.g., Yu \& Tremaine 2002; Elvis et al. 2002; Marconi et al. 2004). The 
latest magnetohydrodynamic (MHD) simulations for disc accretion indicate 
an $\epsilon$ higher than the oft-quoted value of $\sim 0.1$ (e.g., Gammie, 
Shapiro \& McKinney 2004). A higher value of $\epsilon$ tends to increase 
the Salpeter timescale $t_{\hbox{\tiny Sal}}$ and thus makes the mass 
growth of a SMBH via gas accretion more difficult within a short time 
(e.g., Shapiro 2005). Given an estimated age of $\sim 0.9$ Gyr for the 
quasar SDSS 1148+5251 at $z=6.43$, it would not be easy to assemble a 
SMBH of mass $\sim 3\times 10^9\ M_\odot$ from a $\sim 10-100\ M_\odot$ 
seed BH (e.g., the remnant of an imploding core of a massive Population 
III star; Arnett 1996; Heger \& Woosley 2002) even for a sustained 
accretion of baryon matter at the Eddington limit all the time.

While speculative to various extents, possible theoretical resolutions
to this dilemma of rapid SMBH growth in the early Universe include:
(1) more massive seed BHs either from collapses of supermassive
stars (e.g., Shapiro 2004) or from accretion of low angular
momentum baryon materials in the early Universe (e.g.,
Koushiappas, Bullock \& Dekel 2004);
(2) more frequent BH merging processes (e.g., Yoo \&
Miralda-Escud\'e 2004; Shapiro 2005; but see Haiman 2004);
(3) rapid mass growths via a sustained super-Eddington accretion
(e.g., Ruszkowski \& Begelman 2003; Volonteri \& Rees %Fabian
2005). All these proposals with various assumptions might
produce the required mass $\gsim 10^9M_{\odot}$ of a SMBH
at $z=6.43$ through a baryon accretion alone.

Alternatively, a sustained accretion of self-interacting dark matter
(SIDM) particles (e.g., Spergel \& Steinhardt 2000) onto a seed BH
have been modelled to reproduce the observed $M_{\hbox{\tiny BH}}
-\sigma_b$ relation (Ostriker 2000; Hennawi \& Ostriker 2002; cf.
MacMillan \& Henriksen 2002 for an alternative approach). As a different
application of these ideas and as a theoretical contest, here we
entertain the possibility that a proper combination of SIDM and baryon 
accretion at distinct stages might lead to desired features of forming 
SMBHs in the early Universe. It is natural and sensible to imagine that 
on large scales, SIDM particles and baryons are intermingled and mediated 
by gravitational interactions through fluctuations in the early Universe. 
Based on the theoretical knowledge of accreting baryon matter, we 
therefore advance in this paper a two-phase scenario involving accretion 
of both quasi-spherical SIDMs and baryon matter. In \S 2, we describe 
and elaborate our two-phase accretion model scenario in specifics. 
Summary and discussion are contained in \S 3.

\section{The Two-Phase Accretion Scenario}

In our two-phase accretion model for SMBH formation, the first phase is 
featured by a sustained, rapid quasi-spherical and quasi-steady Bondi 
accretion of mainly SIDM particles (a small fraction of baryon matter 
mixed therein) onto a seed BH created at $z\lsim 30$ presumably by a 
core implosion inside a first-generation massive star of Pop III until
reaching a BH mass of $M_{\hbox{\tiny BH}}\sim 10^6\ M_\odot$ during the 
redshift range of $z\sim 20-15$. The second phase of subsequent BH mass 
growth is primarily characterized by a baryon accretion to eventually 
assemble a SMBH of mass $M_{\hbox{\tiny BH}}\sim 10^9\ M_\odot$; such 
SMBHs may form either around $z\sim 6$ for a sustained baryon accretion 
at the Eddington limit or later at lower $z$ for average accretion rates 
below the Eddington limit. For conceptual clarity,
we consider these two major phases separately. On the theoretical
ground, the first phase should gradually evolve into a diffusively
limited accretion of SIDM particles continuing towards the BH. As
time goes on, the initial mixture of SIDM particles and baryon
matter will eventually become more or less separated during the
accretion process in the sense that radiative baryon matter
gradually flatten to a disc accretion which eventually overwhelms
the steady accretion of SIDM particles fading into the diffusively
limited process.

\subsection{Phase I: a sustained spherical accretion 
of mainly SIDM particles onto a seed black hole }

We presume that the initial seed BHs were created by core
implosions of massive Pop III stars with typical remnant BH masses 
of $M_0\sim 10-100M_\odot$ in the redshift range $z\sim 30-10$ [e.g., 
{\it Wilkinson Microwave Anisotropy Probe (WMAP)} observations for 
the excess power in cosmic microwave background provide tantalizing
evidence for the reionization era; see Kogut et al. 2003]; frequent
coalescences of such seed BHs might possibly happen to produce more 
massive seed BHs around the same epoch or shortly thereafter. When 
such a seed BH happens to immerse in a dark matter (DM) halo of low 
angular momentum, it accretes SIDMs together with a small fraction 
of baryon matter mixed therein. As an optimistic approximation, such 
a SIDM accretion is envisaged as grossly spherically symmetric by an 
effective transport of angular momentum outward in the ensemble of
SIDM particles (Ostriker 2000). We define the specific cross section 
as $\sigma_0\equiv\sigma_x/m_x$ for an SIDM particle with a mass $m_x$ 
and a cross section $\sigma_x$; the mean free path is therefore
$\lambda=1/(\rho\sigma_0)$ with $\rho$ being the mass density of the 
SIDM including a small mass fraction of baryon matter. For regions 
within the radial range $r\gsim\lambda$, the SIDM is sufficiently 
dense and may be grossly perceived as a `fluid' (e.g., Peebles 2000; 
Subramanian 2000; Moore et al. 2000; Hennawi \& Ostriker 2002).

As a classical reference of estimates, we begin with the
well-known stationary Bondi (1952) accretion. Given a singular
isothermal sphere (SIS) mass density profile\footnote{While
being highly speculative for a SIDM halo, the SIS density
profile for SIDM and gas was also considered earlier by
%assumed in modelling the $M_{\hbox{\tiny BH}}-\sigma_b$ relation
Ostriker (2000) and King (2003). Ostriker discussed consequences
of other density profiles of $r^{-1}$ (e.g., Navarro, Frenk, \&
White 1995) and $r^{-3/2}$ (e.g., Jing \& Suto 2000; Subramanian,
Cen, \& Ostriker 2000; Lou \& Shen 2004; Bian \& Lou 2005; Yu \& 
Lou 2005). There are other cusped power-law density profiles 
available in the so-called hypervirial family (e.g., Evans \& An 2005). }  
$\rho=C_s^2 /(2\pi Gr^2)$ for a SIDM halo, the mass growth with time 
of a BH embedded in a quasi-spherical SIDM halo is 
$M_{\hbox{\tiny BH}}(t)=2C_s^3t/G$, where $G$ is the gravitational 
constant and $C_s$ is the SIDM halo ``sound speed" (Ostriker 2000). 
Here the ``sound speed" is essentially the local velocity dispersion 
of SIDM particles and is equal to the virial velocity for the SIS 
case. A quantitative comparison of the Bondi accretion with the 
Eddington accretion is given by the ratio
\begin{equation}
\frac{\dot{M}_{\hbox{\tiny Bon}}}{\dot{M}_{\hbox{\tiny Edd}}}
=560\frac{\epsilon}{0.1}\bigg(\frac{C_s}{30\ {\rm km\ s^{-1}}}
\bigg)^3\bigg(\frac{M_{\hbox{\tiny BH}}}{10^6\ M_\odot}\bigg)^{-1}\ ,
\end{equation}
where $\dot{M}_{\hbox{\tiny Bon}}$ and $\dot{M}_{\hbox{\tiny Edd}}$ 
are the Bondi and Eddington mass accretion rates, respectively.
Apparently, given a seed BH mass and a typical halo sound speed (see 
below), the Bondi mass accretion rate $\dot{M}_{\hbox{\tiny Bon}}$ 
dominates over the Eddington mass accretion rate 
$\dot{M}_{\hbox{\tiny Edd}}$. As SIDM particles do not radiate, this 
super-Eddington accretion will proceed without impedence. For sustained 
isothermal spherical self-similar collapses or accretion (e.g., Lou \& 
Shen 2004; Shen \& Lou 2004), the maximum mass growth rate remains in 
the same order of magnitude as that of the steady Bondi accretion.
We emphasize that in the presence of accretion shocks, the central 
mass accretion rate should be modified (Shen \& Lou 2004; Bian \& Lou 2005).

This SIDM accretion phase continues until the mean free path 
$\lambda$ becomes comparable to the accretion radius $r_A$ with 
a corresponding timescale $t_1=\sigma_0C_s/(4\pi G)$, a BH mass
$M_{\hbox{\tiny BH}}(t_1)=\sigma_0 C_s^4/(2\pi G^2)$ and a transitional 
accretion radius $r_A(t_1)=\sigma_0C_s^2/(2\pi G)$. For typical 
parameters, we have the following quantitative estimates
\begin{equation}\label{rf1}
\begin{split}
%&t_1\backsimeq 7.6\times 10^3\hbox{ yr}\ (\sigma_0/0.02\
%\hbox{cm$^2$ g$^{-1}$})(C_s/100\ \hbox{km
%s$^{-1}$})\ ,\\
&t_1\backsimeq 1.1\times 10^5\hbox{ yr}\
(C_s/30\ \hbox{km s$^{-1}$})\
\sigma_0/(1\ \hbox{cm$^2$ g$^{-1}$)}\ ,\\
%&M_{\hbox{\tiny BH},\ t_1}\backsimeq 3.6\times 10^6\ M_\odot\
%(\sigma_0/0.02\ \hbox{cm$^2$ g$^{-1}$})(C_s/100\ \hbox{km
%s$^{-1}$})^4\
%,\\
&M_{\hbox{\tiny BH}, t_1}\backsimeq 1.4\times 10^6M_\odot
(C_s/30\ \hbox{km s$^{-1}$})^4
\sigma_0/(1\ \hbox{cm$^2$ g$^{-1}$})
\ ,\\
%&r_{A,\ t_1}\backsimeq 1.5\ \hbox{pc}\ (\sigma_0/0.02\
%\hbox{cm$^2$ g$^{-1}$})(C_s/100\ \hbox{km s$^{-1}$})^2\ .\\
&r_{A, t_1}\backsimeq 7\ \hbox{pc}\
(C_s/30\ \hbox{km s$^{-1}$})^2
\sigma_0/(1\ \hbox{cm$^2$g$^{-1}$})
\ .
\end{split}
\end{equation}

\begin{figure}
\mbox{\epsfig{figure=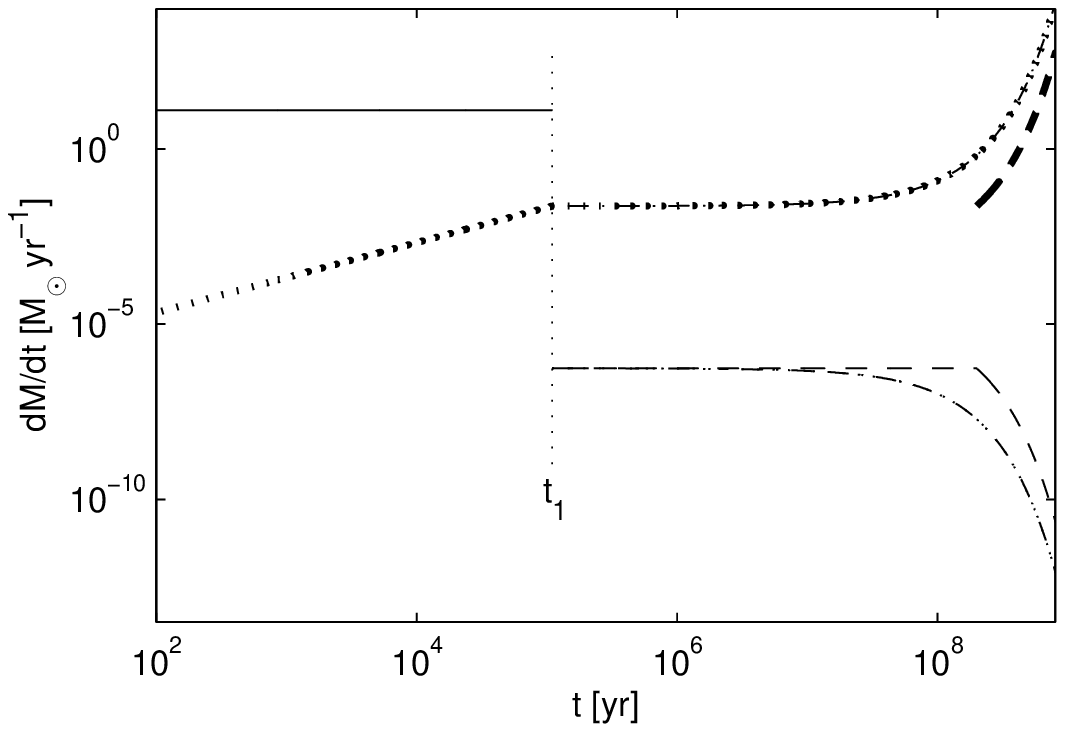,width=\linewidth,clip=}}
\mbox{\epsfig{figure=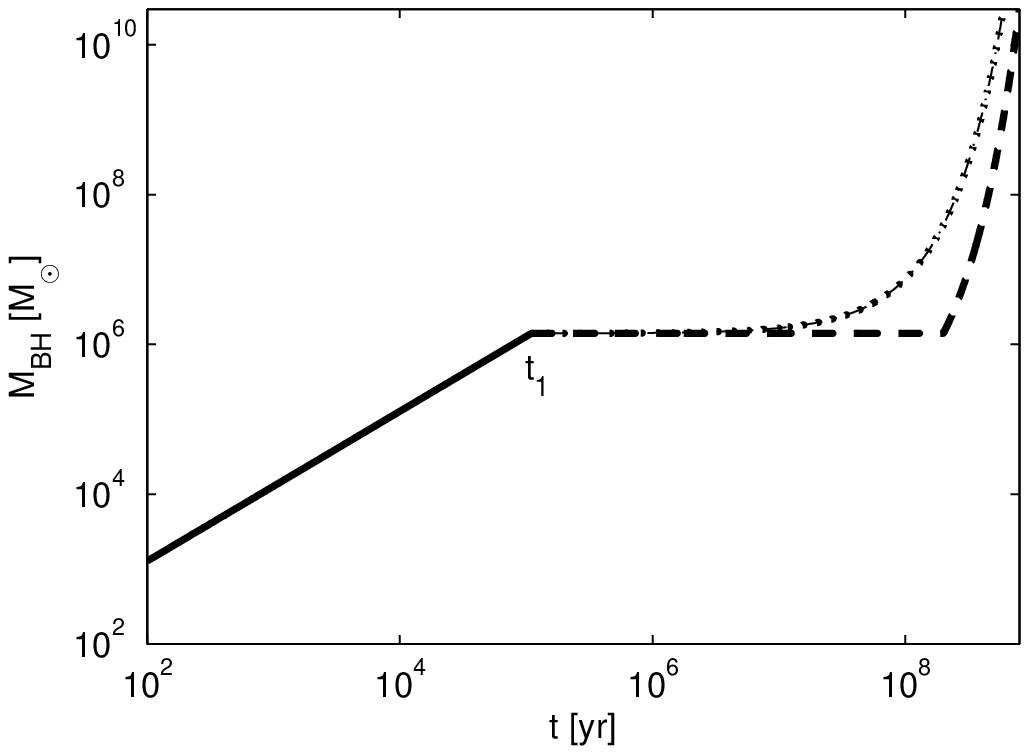,width=\linewidth,clip=}}
\caption{The mass growth history of a black hole in the SIDM halo
mixed with a fraction of baryon matter. The upper and lower panels
are for the histories of mass accretion rate $dM/dt$ and of black hole
mass $M_{\hbox{\tiny BH}}$ growth, respectively. The thin solid line
before $t_1$ represents the Phase I steady Bondi accretion of SIDM 
particles, while the other thin lines are the diffusive limited SIDM 
accretion after $t_1$. The curves of thicker (boldface) linetypes are 
the Phase II baryon accretion at the Eddington rate: here, we consider 
three different cases of the moment that Phase II accretion begins, 
that is, before (dotted), at (dash-dotted) and after (dashed) $t_1$,
respectively. Please note that the dash-dotted line and dotted line 
almost coincide to the right side of $t_1$. The initial seed black 
hole mass due to a Pop III star is $30\ M_\odot$ at $z=20$. The 
input parameters are $C_s=30$ ${\rm km\ s^{-1}}$, $\sigma_0=1.0\ 
{\rm cm^2\ g^{-1}}$ and $\epsilon=0.15$.}\label{fig1}
\end{figure}

For a virialized SIDM halo at high $z$, the typical halo virial 
velocity or local velocity dispersion of SIDM particles 
(mimicked as a `sound speed') may be estimated by 
\begin{equation}\label{rf2}
C_s=8.2(M/10^9M_\odot)^{1/3}(1+z)^{1/2}\ {\rm km\ s^{-1}}\ 
\end{equation}
(e.g., Barkana \& Loeb 2001). Therefore, an estimate of 
$C_s\sim 30\hbox{ km s}^{-1}$ for a virilized SIDM halo 
of mass $M\sim10^9M_\odot$ during $z\sim 20-15$ appears 
plausible. The comoving halo number density $n(M,z)$ with 
mass $M$ at a given $z$ can be calculated from the standard 
hierarchical structure formation model. We adopt an input 
power spectrum computed by Eisenstein \& Hu (1999). For 
cosmological parameters, we take $\Omega_{\rm m}=0.3$,
$\Omega_{\Lambda}=0.7$, $\Omega_{\rm b} =0.045$, $h_0=0.7$,
$\sigma_8=0.9$ and $n=1$ in our model calculations.

Estimated constraints on $\sigma_0$ from both observations and theories 
are briefly summarized at this point. Wandelt et al. (2000) evaluated 
the constraints on $\sigma_0$ and found a $\sigma_0$ range of 
$\sim 0.5-6$ cm$^2$ g$^{-1}$. Yoshida et al. (2000) numerically 
simulated the evolution of a galaxy cluster for $\sigma_0=10, 1$, and 
0.1 cm$^2$ g$^{-1}$, and obtained corresponding radial mass profiles. 
Using the high-resolution X-ray data of {\it Chandra} satellite and 
the assumption of a hydrostatic equilibrium, Arabadjis et al. (2002) 
derived a mass profile for the galaxy cluster MS 1358+62 that peaks 
strongly in the central region. From a comparison with the result of 
Yoshida et al. (2000), they concluded that $\sigma_0\lsim 0.1$ cm$^2$ 
g$^{-1}$. However, Markevitch et al. (2004) pointed out that there 
are certain difficulties with this stringent limit and they provided 
a less stringent limit of $\sigma_0\lsim 1$ cm$^2$ g$^{-1}$.

In general, $\sigma_0$ may well depend on relative velocity 
$v$ of SIDM particles (e.g., Firmani et al. 2000). Here, we 
tentatively adopt a $\sigma_0$ in the form of 
\begin{equation}\label{rf3}
\sigma_0=\bigg(\frac{30\ {\rm km}
\ {\rm s}^{-1}}{v}\bigg)\ {\rm cm}^2\ {\rm g}^{-1},
\end{equation}
where $v$ may be estimated by the isothermal 
`sound speed' $C_s$ parameter.

Around and after the time $t_1$, a transition to a slower diffusively 
limited accretion of SIDM particles will gradually occur. The accretion 
rate is then determined by the rate at which SIDM particles are scattered 
into the loss cone. Meanwhile, the embedded or intermingled baryon matter 
become more concentrated and flattened through radiative processes, and 
their accretion rate towards the central black hole becomes more and 
more important as the diffusively limited accretion of SIDM particles 
approaches a quasi-steady state. The BH continues to accrete SIDM 
particles further but at a much slower rate than the Eddington mass 
accretion rate of baryons as shown by the mass accretion rate ratio 
below from equation (5) of Ostriker (2000) 
\begin{equation}
\frac{\dot{M}_{\hbox{\tiny DM}}}
{\dot{M}_{\hbox{\tiny Edd}}}=
3.4\times 10^{-5}\frac{\epsilon}{0.1}
\frac{\sigma_0}{1\ {\rm cm^2\ g^{-1}}}
\bigg(\frac{C_s}{30\ {\rm km\ s^{-1}}}\bigg)^9
\bigg(\frac{M_{\hbox{\tiny BH}}}{10^6 M_\odot}\bigg)^{-2}.
\end{equation}
For order-of-magnitude estimates, we may safely ignore the SIDM
accretion more or less after the formation of a significantly
enlarged seed BH around the end of Phase I accretion. In this
sense and in reference to the very initial seed BH, we regard it 
as a `secondary seed BH' for the Phase II accretion of baryon 
matter. In Figure \ref{fig1}, we have explored three different 
onset times for the Eddington accretion rate of baryon matter
for comparison. Within our scenario, it is more sensible to have 
the baryon accretion all along with the SIDM accretion, roughly 
corresponding to the heavy dotted curve.

\subsection{Phase II: a disk accretion of baryon matter}

In contrast to baryon matter accretion at the Eddington limit as 
estimated by equation (2), a rapid quasi-spherical accretion of SIDM 
particles during Phase I dominated during the first $\sim 10^5$ yr or 
so and the BH mass rapidly grows to
$M_{\hbox{\tiny BH},\ t_1}\sim 10^6\ M_\odot$ to become a secondary
seed BH for the subsequent baryon matter accretion. After this almost 
instantaneous Phase-I accretion in reference to the Salpeter timescale 
$t_{\tiny\hbox{Sal}}$, the accumulation of SIDMs proceeds at a much 
slower pace with an inefficient loss cone accretion (Ostriker 2000; 
Hennawi \& Ostriker 2002). With favourable environs or sustained
reservoirs of fuels, accretion of baryon matter gradually become dominant 
to increase the BH mass by a factor of $\sim 10^3$ within subsequent 
several Salpeter times [see equation (2)] at the Eddington accretion rate. 
It is this subsequent accretion of baryon matter that eventually assembles 
most 
of the SMBH mass, consistent with observations that the BH mass densities 
in nearby galactic bulges and in active SMBHs are comparable to the mass 
density accreted during the optically bright/obscured QSO phase (e.g., Yu 
\& Tremaine 2002; Fabian 2004; Haehnelt 2004). Based on the explorations 
in Figure 1, it would not matter that much as for the exact moment when 
the Phase II Eddington baryon accretion sets in. In fact, it may begin 
at any moment (either before or after $t_1$). However, the Phase I SIDM 
accretion can produce a massive enough seed BH for the Phase II baryon 
accretion.

Figure 1 illustrates a black hole growth history including the two
phases. The Phase I SIDM accretion increases the black hole mass
substantially, and then the Phase II baryon accretion enhance the
enlarged seed black hole further. The accretion due to SIDM particles 
after $t_1$ peters out rapidly into the diffusively limited regime.

\subsection{Model applications to high-$z$ quasars}

\begin{figure}
\mbox{\epsfig{figure=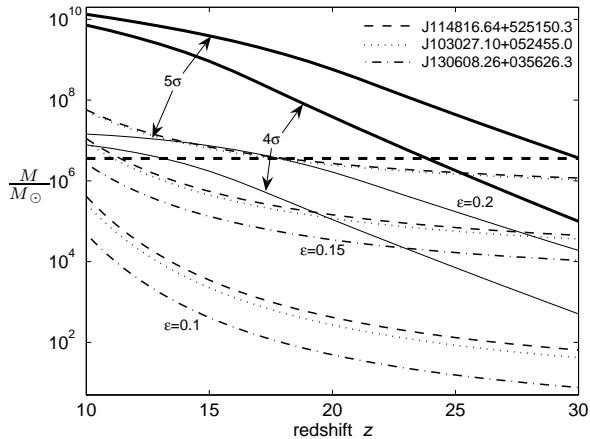,width=\linewidth,clip=}}
\caption{Three sets of dashed, dotted, dash-dotted curves with
radiative efficiency $\epsilon=0.2,\ 0.15,\ 0.1$ respectively are
the required minimum BH masses for $10<z<30$ after the phase-I SIDM
accretion, with distinct line types referring to three reported
high$-z$ quasars (SMBHs). The heavy solid lines show the virialized
dark matter halo mass $M_{\rm{vir}}(z)$ with $4\sigma$ and $5\sigma$ 
fluctuations at different $z$ values. The light solid lines show the 
secondary seed BH mass $M_{\hbox{\tiny BH},\ t_1}$ created by the 
first rapid accretion phase at various $z$ for $C_s$ and $\sigma_0$ 
according to equations (\ref{rf2}) and (\ref{rf3}). The heavy dashed 
line shows $M_{\hbox{\tiny BH},\ t_1}$ with a constant sound speed 
$C_s=100\rm{\ km\ s^{-1}}$ and a smaller cross section 
$\sigma_0=0.02\ \rm{cm^2\ g^{-1}}$.}\label{fig2}
\end{figure}

Figure 2 shows the mass range of forming high$-z$ SMBHs, where we 
take on three observed SDSS high$-z$ quasars with reported SMBH masses
(Fan et al. 2001, 2003; Willott et al. 2003; Vestergaard 2004):
J114816.64+525150.3 ($z=6.43$, $M_{\hbox{\tiny BH}}\sim 3\times
10^9\ M_\odot$), J103027.10+052455.0 ($z=6.28$, $M_{\hbox{\tiny
BH}}\sim 3.6 \times 10^9\ M_\odot$), and J130608.26+035626.3
($z=5.99$, $M_{\hbox{\tiny BH}}\sim 2.4\times 10^9\ M_\odot$)
distinguished by dashed, dotted, and dash-dotted curves,
respectively. Based on our model scenario, we trace BH masses back
to higher $z$, assuming a sustained baryon Eddington accretion and 
three different radiative efficiencies $\epsilon=0.1,\ 0.15,\ 0.2$
to compute the required minimum BH mass of the phase-I accretion as 
a function of redshift $z$. We plot the actual final BH mass 
$M_{\hbox{\tiny BH}, \ t_1}$ of the phase-I accretion (light solid 
curves in Fig. 2) in DM halos with 4-$\sigma$ and 5-$\sigma$ 
fluctuations at a given $z$, using equation (\ref{rf1}) and $C_s$ 
given by equation (\ref{rf2}). As an exploration, we also calculate 
$M_{\hbox{\tiny BH},\ t_1}$ with $C_s\sim 100\hbox{ km s}^{-1}$ 
yet with a smaller specific cross section 
$\sigma_0=0.02\hbox{ cm}^2\hbox{ g}^{-1}$ 
shown by the heavy dashed line in Fig. 2.

As shown in Fig. 2, our model (i.e., light solid curves) can readily
account for the three high$-z$ SMBHs for $\epsilon=0.1$ and $0.15$,
without invoking either super-Eddington baryon accretion or numerous 
BH merging processes. For a higher $\epsilon=0.2$, it is unnatural to 
explain the presence of high-$z$ SMBHs with 4-$\sigma$ fluctuations. 
For 5$-\sigma$ fluctuations, the mass requirement can be just met. It 
is clear that only under rare circumstances, extremely massive SIDM 
halos may give birth to seed BHs with required Phase-I masses in our 
scenario.

There are fundamental differences between the two main accretion 
phases, among which the most important one is that baryon accretion
produces strong detectable feedback (e.g., intense radiations and
outflows or jets etc) into surroundings. Reviving some ideas of Silk
\& Rees (1998), King (2003) found a simple yet remarkable association
between accretion and outflows which plausibly leads to the local
$M_{\hbox{\tiny BH}}-\sigma_b$ relation (e.g., Tremaine et al. 2002)
at the termination of accreting a SMBH. The phase-II accretion in
our scenario involves processes largely similar to those outlined by
King (2003), including Eddington luminosity, baryon accretion,
intense outflows, and so forth; an $M_{\hbox{\tiny BH}}-\sigma_b$
relation (see equation 15 of King 2003) can be established as the
baryon accretion goes on and the SMBH is continuously assembled. 
As described, such an $M_{\hbox{\tiny BH}}-\sigma_b$ relation may 
not emerge before ending the phase-I accretion. To be specific, we 
combine equations (14) and (7) of King (2003) as
\begin{equation}
M_{\hbox{\tiny BH}}=\frac{\kappa f_g\sigma_b^4}{2\pi
G^2}\bigg(\frac{v_m}{\sigma_b}\bigg)^2\backsimeq
1.5\times10^8\sigma_{b200}^4(v_m/\sigma_b)^2\ M_\odot\ ,
\end{equation}
where $\kappa$ is the electron scattering opacity, $f_g\backsimeq 0.16$ 
is the gas fraction, $v_m$ is the mass shell velocity driven by outflows 
and $\sigma_{b200}$ is the bulge stellar velocity dispersion $\sigma_b$ 
in unit of $200\hbox{ km s}^{-1}$. Potentially, it would also be very 
important to incorporate effects of magnetic field in the baryon 
accretion phase (Yu \& Lou 2005; Wang \& Lou 2005 in preparation). 
During the accretion phase, $v_m<\sigma_b$ at first and gradually $v_m$ 
approaches $\sim\sigma_b$ as the accretion diminishes; meanwhile, the 
$M_{\hbox{\tiny BH}}-\sigma_b$ relation appears. During the QSO/AGN 
phase, BH masses may drop below the $M_{\hbox{\tiny BHu}}-\sigma_b$ 
relation for normal galaxies, as indicated by some results of 
narrow-line Seyfert 1 galaxies (e.g., Bian \& Zhao 2004; Grupe \& 
Mathur 2004). Further observations and data analysis of BH masses 
versus velocity dispersions are needed to explore these aspects.

On the basis of our scenario, we may accommodate the high$-z$ and 
low$-z$ quasars in a unified framework: There exists a distribution of 
baryon accretion rates onto secondary seed BHs of $\sim 10^6\ M_\odot$ 
produced between $z\sim 20-15$ by rapidly accreting SIDMs during Phase
I. Given specific situations, such secondary seed BHs may continue to 
accrete baryon matter with high rates
%(near the Eddington limit and stay in the ``high" 
%state of the S-curve almost continuously)
until they run out of fuels or the accretion rate becomes considerably 
lower than the Eddington limit; this then gives rise to high$-z$ SMBHs 
of $\sim 10^{8-9}\ M_\odot$. Those secondary seed BHs with lower rates 
of accreting baryon matter will evolve into SMBHs much later as already 
discussed above. Given the paucity of high$-z$ quasars detected so far, 
our scenario implies that BH systems with high accretion rates should 
be extremely rare.

\section{Summary and Discussion}

We summarize the two-phase accretion model of SMBH formation 
in the early Universe below and discuss a few observational 
implications. 

Phase I involves a rapid quasi-spherical accretion of mainly SIDM 
particles mixed with a small fraction of baryon matter onto an 
initial seed BH created by a Pop III star. Such a BH grows rapidly
to $\sim 10^6\ M_\odot$ within a fairly short timescale of 
$\sim 10^5$ yr.

Phase II involves an accretion of primarily baryon matter (initially 
mixed with SIDM particles) at the Eddington limit to accumulate 
most of the BH mass. Since the BH mass is sufficiently massive at 
the beginning of this phase II, it takes only several Salpeter 
$e-$folding times to grow a BH mass of $\sim 10^9\ M_\odot$ 
according to the estimate of an exponential growth by equation (2).

The transition from Phase I to the much slower diffusion limited 
accretion of SIDM particles goes on concurrently with the gradual 
dominance of baryon accretion at the Eddington limit in Phase II.

The Phase-I accretion of SIDMs is crucial to produce a sufficiently
massive secondary seed BH for further growth by accreting baryon matter. 
In this scenario, the reported high$-z$ SMBHs of $\sim 10^9\ M_\odot$ can 
be produced, without invoking the hypotheses of either super-Eddington 
baryon accretion or extremely frequent BH merging processes. Based on a 
sample of six quasars with $z>5.7$ observed by the SDSS, Fan et al. 
(2003) estimated a comoving density of such luminous quasars at 
$z\sim 6$ and found these quasars showing a $\sim 5\sigma$ peak in 
the density field. This inference may be readily accounted for by 
our cosmological model results shown in Fig. 2. We attempt to combine 
the models of quasi-spherical and quasi-steady SIDM accretion with a 
baryon accretion to give more plausible schemes of accretion leading 
to early formation of SMBHs with masses $\sim 10^9M_{\odot}$ at high 
redshifts of $z\gsim 5-6$.

We now elaborate on consequences of this two-phase scenario. If most 
SMBHs form by this two-phase accretion with fairly rare BH mergers, 
the currently observed upper mass limit for a SMBH $\lsim 10^{10}\ 
M_\odot$ constrains the combination of the specific SIDM cross 
section $\sigma_0$ and the effective baryon accretion time $t_2$ 
(cf. equation 4), such that
%\footnote{This total baryon accretion time $t_2$ can 
%be considerably larger than the QSO lifetime $t_Q$.}
%\footnote{Should we take into account of BH merging
%processes by an amplification of $\sim 10^4$ (e.g.,
%Yoo \& Miralda-Escud\'e 2004), the upper limit would
%be reduced by a factor of $\sim 10^4$. }
\begin{equation}\label{rf4}
\begin{split}
\frac{\sigma_0}{1\hbox{ cm$^2$\ g$^{-1}$}}
\bigg(\frac{C_s}{100\hbox{ km s}^{-1}}\bigg)^4
\exp(t_2/t_{\hbox{\tiny Sal}})\lsim 60\ ,
\end{split}
\end{equation}
where the Eddington luminosity is adopted.\footnote{Should we 
take into account of BH merging processes by a significant mass 
amplification factor of $\sim 10^4$ (e.g., Yoo \& Miralda-Escud\'e 
2004), then this upper limit would be reduced by a factor of 
$\sim 10^4$ accordingly.} For SIDM halos formed at low $z$, they may 
become massive enough to make the virialized velocity dispersion 
or `sound speed' of the order of $C_s\sim 100\hbox{ km s}^{-1}$. 
So the baryon accretion time $t_2$ cannot exceed $\sim 
5t_{\hbox{\tiny Sal}}$ for low$-z$ QSOs. For typical parameters 
of $\epsilon=0.1$ and $L=L_{\hbox{\tiny Edd}}$, it would require 
a $t_2<2\times 10^8$ yr, consistent with the observational QSO 
lifetime of $t_Q\simeq 10^7-10^8$ yr estimated for low$-z$ QSO 
populations (e.g., Yu \& Tremaine 2002; McLure \& Dunlop 2004).

If we take on reference values of equation (\ref{rf1}) for the 
particular $z=6.43$ SMBH with $\sigma_0=1\ \rm{cm^2\ g^{-1}}$ and
$C_s=30\hbox{ km s}^{-1}$, it would have accreted baryon matter over
$\sim 8t_{\hbox{\tiny Sal}}$ ($\sim 3\times 10^8\hbox{ yr}$) for
$\epsilon=0.1$ and $L=L_{\hbox{\tiny Edd}}$, in the absence of
accretions at super-Eddington rates and of BH merging processes.
Such a Phase II accretion time is much longer than the low$-z$ QSO 
lifetime, but accounts for only $\sim 1/3$ of ages for these quasars. 
We speculate that high$-z$ quasars may have longer effective accretion 
times than low$-z$ quasars as the former might have more abundant fuel 
supplies. The plausible physical reasons include: 
(1) gas materials in galaxies have been much consumed 
by star formation activities in the low$-z$ quasars;
(2) interactions among galaxies might have been more frequent in
the early Universe, that may trigger high accretion activities.
Insufficient or interrupted baryon accretion would very likely 
leave behind less massive SMBHs with masses well below the
$M_{\hbox{\tiny BH}}-\sigma_b$ relation curve.

%On the contrary, if the duration of the second accretion phase is
%limited by $t_Q\lsim 5\times10^8\hbox{ yr}$, then the $z=6.43$
%SMBH cannot grow to such mass unless $\sigma_0\gtrsim 0.2\hbox{
%cm$^2$\ g$^{-1}$}$.

%An interesting question arises: {\it how early and how massive
%can an SMBH form?} Future optical/X-ray ultra-deep surveys are
%very likely to reveal SMBH at even higher redshift.

We note that the Phase I accretion will not affect observations
at low$-z$ galaxies. First, the Phase I SIDM Bondi accretion is 
more dominant than the Eddington accretion of baryons only for 
very early ($z>10$) SIDM halos, as such halos tend to have a high 
`sound speed' $C_s$ (cf. equations $3-5$). Secondly, the phase-I 
accretion might be severely limited by the inner density profile 
of the SIDM halo. We have used a SIS mass density profile 
$\rho\propto r^{-2}$. For a self-similar accretion at a given time 
$t$, one would have $\rho\propto r^{-3/2}$ (e.g., Jing \& Suto 2000; 
Lou \& Shen 2004; Bian \& Lou 2005; Yu \& Lou 2005). Numerical 
N-body simulations (e.g., Navarro et al. 1997) indicate an inner 
$\rho\propto r^{-1}$, i.e. the NFW profile (see discussions of Ostriker 
2000). For the more inclusive family of the hypervirial models, we have 
$\rho\propto r^{-(2-p)}$ with the index $p$ falling in the range of
$0<p\leq 2$ (Evans \& An 2005). A shallower profile may lead to a 
less efficient accretion of SIDMs (e.g., Hennawi \& Ostriker 2002). 
Observations find that the inner density profiles (e.g., a 
$\rho\propto r^{-0.5}$) of dark matter halo of nearby galaxies are 
shallower than both the SIS and NFW density profiles, and therefore 
the resulting secondary seed BH mass $M_{\hbox{\tiny BH},\ t_1}$ may 
be expected to be smaller.

Specific to our Milky Way galaxy, the central black hole has been 
inferred to possess a mass of $\sim 4\times 10^{6}M_{\odot}$ by
stellar dynamic diagnostics. For such a less massive SMBH, there 
are many possible ways to assemble such a BH and it is not 
necessary to invoke our two-phase scenario.

Finally, we note possible observational signatures and consequences of 
the two-phase scenario envisaged in this paper. If some large regions in 
the early Universe were somehow distributed with considerable less baryon 
matter (e.g., the gravitational potential well of the halo may not be 
deep enough to bind the high speed baryons due to supernova or hypernova 
feedbacks), then we may find some DM halos with BHs at their centres but 
without forming host galaxies. In other words, these BHs grow entirely 
by Phase I SIDM quasi-spherical Bondi accretions ended with diffusively 
limited phase, and the baryon accretion has never occurred in a 
significant manner. Such unsual systems of DM halo-BHs might be 
revealed directly by chance through gravitational lensing effects.

Very recently, Magain et al. (2005) reported the discovery of a bright 
quasar HE0450-2958 yet without a massive host galaxy. The black hole of 
HE0450-2958 may well be embedded within a dark matter halo (`dark galaxy'), 
constituting a DM halo-BH system. Ambient interactions with such a 
massive object could more readily explain the ring-like starburst in the 
neighbouring galaxy as well as the capture of gas materials, leading to 
the eventual onset of quasar activities we observe. Our model can readily 
account for such kind of phenomena. The phase I SIDM accretion in the 
dark matter halo produced a seed black hole with a mass range of 
$\sim 10^6-10^7\ M_\odot$, but the phase II baryon accretion has never 
occurred due to the absence of a host baryon galaxy. As such a massive 
seed black hole travelled across a neighbouring disc galaxy, it began 
to induce baryon accretion activities violently to give rise to a 
bright quasar. We will further develop models on the basis of such a 
scenario in a separate paper.

Another interesting piece of observational evidence comes from the 
galaxy NGC 4395 (Peterson et al. 2005). A very `small' low-luminosity 
SMBH ($\sim 3\times10^5\ M_\odot$) resides in the bulgeless galaxy 
NGC 4395, implying that the stellar bulge is not a necessary 
prerequisite for a black hole in the nucleus of a galaxy. Our model 
can provide a plausible explanation in the sense that a relatively 
small SMBH is the product of the phase I SIDM accretion, while the 
phase II baryon accretion at the Eddington rate is almost absent in 
a bulgeless galaxy (of course, a very low rate baryon accretion might 
take place).

\section*{Acknowledgments}
For JH and YS, this research was inspired by {\it Black Hole
Accretion} lectures by J.F. Lu in December 2004. For YQL, this 
research was supported in part by the ASCI Center for Astrophysical 
Thermonuclear Flashes at the Univ. Chicago under Department of 
Energy contract B341495, by the Special Funds for Major State Basic 
Science Research Projects of China, by the THCA, by the Collaborative 
Research Fund from the National Natural Science Foundation of China 
(NSFC) for Young Outstanding Overseas Chinese Scholars (NSFC 10028306)
at the National Astronomical Observatories of China, Chinese Academy 
of Sciences, by NSFC grants 10373009 and 10533020 (YQL) at Tsinghua 
University, and by the Yangtze Endowment from the Ministry of Education 
at Tsinghua University. The hospitality and support of the Mullard Space 
Science Laboratory at University College London, U.K., of School of 
Physics and Astronomy, University of St Andrews, Scotland, U.K., and 
of Centre de Physique des Particules de Marseille (CPPM/IN2P3/CNRS) 
et Universit\'e de la M\'editerran\'ee Aix-Marseille II, France are 
also gratefully acknowledged. Affiliated institutions of YQL share 
this contribution. SNZ acknowledges support by NSFC grant 10233030 
and by NASA's Marshall Space Flight Center and through NASA's Long 
Term Space Astrophysics Program.

%\newpage


\begin{thebibliography}{99}

\bibitem{Arabadjis} Arabadjis J. S., Bautz M. W.,
Garmire G. P., 2002, ApJ, 572, 66

\bibitem{Arnett96} Arnett D., 1996, Supernovae and
Nucleosynthesis (Princeton: Princeton University Press)
%implosion of a 100solar mass star to leave a black
%hole of 1/4 initial mass (i.e., 25 solar mass)

\bibitem{Bardeen} Bardeen J. M., 1970, Nature, 226, 64
%Kerr Metric Black Holes

\bibitem{Barkana} Barkana R., Loeb A., 2001, Phys. Rep., 349, 125

\bibitem{BianLou} Bian F.Y., Lou Y.-Q., 2005, 
MNRAS, in press (astro-ph/0509137)
%Spherical Isothermal Self-Similar Shock Flows

\bibitem{Bondi} Bondi H., 1952, MNRAS, 112, 195
%Bondi accretion

%\bibitem{Carlsonetal} Carlson E. D., Machacek M. E.,
%Hall L. J., 1992, ApJ, 398, 43
%--52; related collisional dark matter

\bibitem{Eisenstein} Eisenstein D., Hu W., 1999, ApJ, 511, 5
%Power Spectra of CDM

\bibitem{Elvis} Elvis M., Risaliti G., Zamorani G., 2002, ApJ, 565, L75
%--L77;
%Most Supermassive Black Holes Must Be Rapidly Rotating

\bibitem{EvansAn05} Evans N. W., An J., 2005, MNRAS, 360, 492
%-498.
%Hypervirial models of stellar systems

\bibitem{Fabian} Fabian A. C., 2004,
%Carnegie Obs. Astrophys. Ser. 1,
in Coevolution of Black Holes and Galaxies, ed.
L. C. Ho, (Cambridge: Cambridge Univ. Press), p447
%Obscured Active Galactic Nuclei and
%Obscured Accretion X-ray observations

\bibitem{Fan01} Fan X. et al., 2001, AJ, 122, 2833
%A Survey of z>5.8 Quasars in the Sloan Digital Sky Survey.
%I. Discovery of Three New Quasars and the Spatial Density
%of Luminous Quasars at z\~6

\bibitem{Fan03} Fan X. et al., 2003, AJ, 125, 1649
%A Survey of z>5.7 Quasars in the Sloan Digital Sky Survey.
%II. Discovery of Three Additional Quasars at z>6

%\bibitem{Fan04} Fan X. et al., 2004, AJ, 128, 515

\bibitem{Ferrarese} Ferrarese L., Merritt D., 2000, ApJ, 539, L9
%$M_{BH}-\sigma_b$ relation for local galaxies, index=4.8\pm 0.5

\bibitem{Firmani} Firmani C. et al., 2000, MNRAS, 315, L29
%D'Onghia, E., Avila-Reese, V., Chincarini, G., \& Hern¨¢ndez, X.
%Evidence of self-interacting cold dark matter
%from galactic to galaxy cluster scales

\bibitem{Gammie} Gammie C. F., Shapiro S. L.,
McKinney J. C., 2004, ApJ, 602, 312
%Black Hole Spin Evolution MHD disk simulations

\bibitem{Gebhardt} Gebhardt K. et al., 2000, ApJ, 539, L13
%$M_{BH}-\sigma_b$ relation for local galaxies, index=3.75\pm 0.3
%15 authors;
%Black Hole Mass Estimates from Reverberation
%Mapping and from Spatially Resolved Kinematics

%\bibitem{Grupe} Grupe D., Mathur S., 2004, ApJ, 606, L41
%M_{BH}-sigma_b relation for a complete sample of soft
%X-ray-selected active galactic nuclei

\bibitem{Haehnelt} Haehnelt M. G., 2004,
%Carnegie Obs. Astrophys. Ser. 1,
in Coevolution of Black Holes and
Galaxies, ed. L. C. Ho, (Cambridge: Cambridge Univ. Press), p406
%Joint Formation of Supermassive Black Holes and Galaxies
%review article

\bibitem{Haiman} Haiman Z., 2004, ApJ, 613, 36
%Constraints from gravitational recoil on the growth
%of SMBHs at high redshift suggest rapid growth by
%super-Eddington accretion  instead of BH mergers

\bibitem{Haring} H\"aring N., Rix H., 2004, ApJ, 604, L89
%On the BH Mass-Bulge Mass Relation
%find a tight relation between M_BH and M_bulge,
%see also Magorrian et al. 1998 for a looser relation

\bibitem{Heger} Heger A., Woosley S. E., 2002, ApJ, 567, 532
%The Nucleosynthetic Signature of Population III

\bibitem{Hennawi} Hennawi J. F., Ostriker J. P., 2002, ApJ, 572, 41
%--54;
%Observational Constraints on the SIDM Scenario
%and the Growth of Supermassive BHs

\bibitem{JingSuto00} Jing Y. P., Suto Y., 2000, ApJ, L69
%--L72;
%The Density Profiles of the Dark Matter Halo Are Not Universal

\bibitem{King} King A., 2003, ApJ, 596, L27
%Black Holes, Galaxy Formation, and the $M_{BH}-\sigma_b$ relation

\bibitem{Kogut} Kogut A. et al., 2003, ApJS, 148, 161
%--173; 11 authors;
%First-Year Wilkinson Microwave Anisotropy Probe (WMAP)
%Observations: Temperature-Polarization Correlation

\bibitem{Kormendy} Kormendy J., Richstone D.,
1995, ARA\&A, 33, 581

\bibitem{KBD} Koushiappas S. M., Bullock J. S.,
Dekel A., 2004, MNRAS, 354, 292
%304;
%Massive black hole seeds from low angular momentum material
%mentioned: experience an efficient Lin-Pringle viscosity
%that transfers angular momentum outwards

\bibitem{Laor} Laor A., 2001, ApJ, 553, 677
%On the Linearity of the Black Hole-Bulge Mass
%Relation in Active and in Nearby Galaxies

\bibitem{LouShen}Lou Y.-Q., Shen Y., 2004, MNRAS, 348, 717
%-734; Envelope Expansion with Core Collapse--
%I. Spherical Isothermal Similarity Solutions

\bibitem{Lynden-Bell69} Lynden-Bell D., 1969, Nature, 223, 690

\bibitem{MacMillan} MacMillan J. D., Henriksen R. N., 2002, ApJ, 569, 83
%BH growth in dark matter and the M_BH-\sigma_b
%relation Cited Ostriker 2000's work

\bibitem{}Magain, P. et al. 2005, Nature, 437, 381
%Discovery of a bright quasar without a massive host galaxy

\bibitem{Magorrian} Magorrian J. et al., 1998, AJ, 115, 2285
%S. Tremaine, D. Richstone, R. Bender, G. Bower, A. Dressler,
%S. M. Faber, K. Gebhardt, R. Green, C. Grillmair, J. Kormendy,
%and T. Lauer;
%The Demography of Massive Dark Objects in Galaxy Centers

\bibitem{Marconi} Marconi A., Risaliti G., Gilli R., Hunt
L. K., Maiolino R., Salvati M., 2004, MNRAS, 351, 169
%--185;
%Local supermassive black holes, relics of active
%galactic nuclei and the X-ray background

\bibitem{Markevitch} Markevitch M. et al., 2004, ApJ, 606, 819
%Gonzalez, A. H., Clowe, D., Vikhlinin, A.,
%Forman, W., Jones, C., Murray, S., Tucker, W.
%Direct Constraints on the Dark Matter Self-Interaction
%Cross Section from the Merging Galaxy Cluster 1E 0657-56

\bibitem{McLure} McLure R. J., Dunlop J. S.,
2004, MNRAS, 352, 1390
%--1404;   astro-ph/0405393
%The cosmological evolution of quasar black hole masses
%conclude L/L_Edd\le 1

\bibitem{Moore} Moore B., Gelato S., Jenkins A.,
Pearce F. R., Quilis V., 2000, ApJ, 535, L21
%Collisional versus Collisionless Dark Matter

\bibitem{Murray} Murray N., Quataert E.,
Thompson T. A., 2005, ApJ, 618, 569
%Norman; Eliot; Todd;
%On the Maximum Luminosity of Galaxies and
%Their Central Black Holes: Feedback from
%Momentum-Driven Winds.

\bibitem{Navarro} Navarro J. F., Frenk C. S.,
White S. D. M., 1997, ApJ, 490, 493
%A universal density profile from hierarchical clustering.
%Astrophy. J., 490, 493-508 (1997). Julio; Carlos; Simon

\bibitem{Ostriker} Ostriker J. P., 2000, Phys. Rev. Lett., 84, 5258
%Collisional Dark Matter and the Origin of Massive Black Holes

\bibitem{Page} Page M. J., Stevens J. A., Mittaz J. P. D.,
Carrera F. J., 2001, Science, 294, 2516
%Submillimeter Evidence for the Coeval Growth
%of Massive Black Holes and Galaxy Bulges

\bibitem{Peebles05} Peebles P. J. E., 2000, ApJ, 534, L127
%-L129;
%Fluid Dark Matter

\bibitem{Peebles05} Peterson B. M., et al., 2005, sumitted to ApJ (astro-ph/0506665)

\bibitem{RB} Ruszkowski M., Begelman M. C., 2003, ApJ, 586, 384
%Eddington Limit and Radiative Transfer in Highly
%Inhomogeneous Atmospheres Super-Eddington luminosity

\bibitem{Salpeter} Salpeter E. E., 1964, ApJ, 140, 796

\bibitem{Shapiro} Shapiro S. L., 2004,
%Carnegie Obs. Astrophys. Ser. 1,
in Coevolution of Black Holes and Galaxies, ed. L. C. Ho,
(Cambridge: Cambridge University Press), p103
%Formation of Supermassive Black Holes: Simulations
%in General Relativity contain review and references
%of supermassive stars

\bibitem{Shapiro} Shapiro S. L., 2005, ApJ, 620, 59
%Spin, Accretion and the Cosmological Growth of SMBHs
%disk accretion of gas will cause the BH quickly to
%rotate at a/M\sim 0.95 and favors the MHD disk
%simulation rather than a standard thin disk model

\bibitem{ShenLou04} Shen Y., Lou Y.-Q., 2004, ApJ, 611, L117
%-L120; Shocked Self-Similar Collapses
%and Flows in Star Formation Processes

\bibitem{Silk} Silk J., Rees M. J., 1998, A\&A, 331, L1
%Quasars and galaxy formation self-regulating
%models, see also King, A. 2003, ApJ, 596, L27

\bibitem{Spergel} Spergel D. N., Steinhardt P. J.,
2000, Phys. Rev. Lett., 84, 3760
%Observational Evidence for Self-Interacting Cold Dark Matter

\bibitem{Subramanian00} Subramanian K., 2000, ApJ, 538, 517
%-527; Kandaswamy
%Self-Similar Collapse and the Structure
%of Dark Matter Halos: A Fluid Approach

\bibitem{SubramanianCenOstriker} Subramanian K.,
Cen R. Y., Ostriker J. P., 2000, ApJ, 538, 528
%-542.
%The Structure of Dark Matter Halos
%in Hierarchical Clustering Theories

%\bibitem{Tegmark} Tegmark M. et al., 2004, PRD, 69, 103501
%SDSS+WMAP cosmological parameters

\bibitem{Tremaine} Tremaine S. et al., 2002, ApJ, 574, 740
%the slope and normalization of the local M_{BH}-\sigma_b
%relation M_{BH}\propto \sigma_b^{4.02}
%total 15 authors.

\bibitem{Vestergaard}Vestergaard M., 2004, ApJ, 601, 676
%Early Growth and Efficient Accretion of
%Massive Black Holes at High Redshift

\bibitem{Volonteri}Volonteri M., Rees M. J., 2005,
submitted to ApJ Letters, astro-ph/0506040
%Rapid growth of high redshift black holes

\bibitem{Walter} Walter F. et al., 2004, ApJ, 615, L17
%Resolved Molecular Gas in A Quasar Host Galaxy at Redshift z=6.42
%violation of the $M_{BH}-\sigma_b$ relation for local galaxies

\bibitem{Wandelt} Wandelt B. D. et al., 2000,
in Proceedings of Dark Matter,
%2000,
astro-ph/0006344
%Self-Interacting Dark Matter

\bibitem{Willott} Willott C. J., McLure R. J.,
Jarvis M. J., 2003, ApJ, 587, L15
%A 3*10^9 $M_\odot$ BH in the Quasar SDSS J1148+5251 at z=6.41

\bibitem{Yoo} Yoo J., Miralda-Escud\'e J., 2004, ApJ, 614, L25
%Formation of the Black Holes in the Highest
%Redshift Quasars Modelling by BH mergers

\bibitem{Yoshida} Yoshida N., Springel V.,
White S. D. M., Tormen G., 2000, ApJ, 544, L87

\bibitem{YuLou} Yu C., Lou Y.-Q., 2005, 
MNRAS, in press (astro-ph/0509242)
%Envelope Expansion with Core Collapse II. Quasi-Spherical 
%Self-Similar Solutions for an Isothermal Magnetofluid

\bibitem{Yu} Yu Q., Tremaine S., 2002, ApJ, 335, 965
%Observational constraints on growth of massive black holes

\end{thebibliography}
\end{document}